\documentclass[preprint,aps,12pt,showpacs,nofootinbib,tightenlines]{revtex4}
\usepackage{amsmath}
\usepackage{amssymb}
\usepackage{epsfig}
\usepackage{graphicx}
\usepackage{subfigure}
\usepackage{color}
\begin{document}
\def\pslash{\rlap{\hspace{0.02cm}/}{p}}
\def\eslash{\rlap{\hspace{0.02cm}/}{e}}
\title {One-loop QCD correction to top pair production in the littlest Higgs model with T-parity at the LHC}
\author{\footnotesize Bingfang Yang}\email{yangbingfang@htu.edu.cn}
\author{ \footnotesize Ning Liu}\email{wlln@mail.ustc.edu.cn}

\affiliation{\footnotesize $^1$Institute of Theoretical Physics,
Henan Normal University, Xinxiang 453007, China
   \vspace*{1.5cm}  }

\begin{abstract}

In this work, we investigate the one-loop QCD correction to top pair
production in the littlest Higgs model with T-parity at the LHC. We
calculate the relative correction of the top pair production cross
section and top-antitop spin correlation at the LHC for
$\sqrt{s}=8,14$ TeV. We find that the relative corrections of top
pair production cross section can reach about $-0.35\%$, and the
top-antitop spin correlation can reach $1.7\%$($2\%$) at the 8(14)
TeV LHC in the favorable parameter space.

\end{abstract}
 \maketitle
\section{ Introduction}
Since the top quark was discovered at the Fermilab Tevatron in
1995\cite{topTeV}, it has always been one of the hottest topics in
particle physics. So far, the top quark is the heaviest particle
discovered, with a mass close to the electroweak symmetry breaking
scale. Thus it is a wonderful probe for the electroweak breaking
mechanism and new physics (NP) beyond the standard model (SM). As a
genuine top quark factory, the LHC will copiously produce the top
quark events and can provide a good opportunity to scrutinize the
top quark properties.

In order to solve the hierarchy problem, the little Higgs model was
proposed \cite{little Higgs}, where the Higgs boson is constructed
as a pseudo-Goldstone boson. The littlest Higgs model (LH)\cite{LH}
provides an economical realization for this theory, however, this
model suffers strong constraints from electroweak precision tests
\cite{constraintLH}. A feasible way to relax this constraints is to
introduce a discrete symmetry called T-parity\cite{T-parity} in the
LH model. This resulting model is referred to as the littlest Higgs
model with T-parity (LHT). The LHT model predicts many new
particles, such as heavy gauge bosons, mirror fermions and heavy top
partners, they can interact with the top quark and contribute to the
top pair production at the loop level.

At the LHC, top quarks can be mostly produced in $t\bar{t}$
production through strong interactions and up to now the $t\bar{t}$
production has been measured in different channels with remarkable
accuracy\cite{toppair}. In general, QCD controls the theoretical
predictions for the $t\bar{t}$ production in both the SM and NP at
hadron colliders, and the QCD high order corrections play a key role
for the accurate theoretical predictions. Therefore, it is necessary
to perform the QCD high order calculations in order to test the SM
and search for NP. In our previous work, we have studied the one
loop electroweak effects on the $t\bar{t}$ production process in the
LHT at the LHC\cite{ewtoplht}. In this paper, we consider the latest
experimental constraints and calculate the one loop QCD corrections
to the $t\bar{t}$ production process in the LHT at the LHC. Since
the new interactions between top quark and LHT particles can not
only affect the $t\bar{t}$ production rate but also the spin
polarization\cite{Cao:2011ew}, we also discuss the LHT corrections
to the spin polarization in the $t\bar{t}$ production process.

This paper is organized as follows. In Sec.II we give a brief review
of the LHT model related to our work. In Sec.III we give a brief
description for the one-loop QCD calculations in the LHT model. In
Sec.IV we show the numerical results and some discussions. Finally,
we make a short summary in Sec.V.

\section{ A brief review of the LHT model}
The LHT model is based on a $SU(5)/SO(5)$ non-linear $\sigma$ model.
At the scale $f\sim\mathcal{O}(TeV)$, the global group $SU(5)$ is
spontaneously broken into $SO(5)$ by a $5\times5$ symmetric tensor
and the gauged subgroup $[SU(2)\times U(1)]^{2}$ of $SU(5)$ is
broken into the SM gauge group $SU(2)_{L}\times U(1)_{Y}$. After the
symmetry breaking, four new heavy gauge bosons
$W_{H}^{\pm},Z_{H},A_{H}$ appear, whose masses up to $\mathcal
O(\upsilon^{2}/f^{2})$ are given by
\begin {equation}
M_{W_{H}}=M_{Z_{H}}=gf(1-\frac{\upsilon^{2}}{8f^{2}}),M_{A_{H}}=\frac{g'f}{\sqrt{5}}
(1-\frac{5\upsilon^{2}}{8f^{2}})
\end {equation}
with $g$ and $g'$ being the SM $SU(2)$ and $U(1)$ gauge couplings,
respectively.

In order to preserve the T-parity, a copy of quarks and leptons with
T-odd quantum number are added. We denote the mirror quarks by
$u_{H}^{i},d_{H}^{i}$, where $i= 1, 2, 3$ are the generation index.
In order to cancel the quadratic divergences to the Higgs boson mass
arising from the SM top quark, an additional T-even top partner
$T^{+}$ that has its associated T-odd mirror quark $T^{-}$ are
introduced. The new fermions which can contribute to the one loop
QCD correction of top quark pair production are
$u_{H}^{i},d_{H}^{i},T^{+},T^{-}$, whose masses up to $\mathcal
O(\upsilon^{2}/f^{2})$ are given by
\begin{eqnarray}
&&m_{d_{H}^{i}}=\sqrt{2}\kappa_if\\
&&m_{u_{H}^{i}}=m_{d_{H}^{i}}(1-\frac{\upsilon^2}{8f^2})\\
&&m_{T^{+}}=\frac{f}{v}\frac{m_{t}}{\sqrt{x_{L}(1-x_{L})}}[1+\frac{v^{2}}{f^{2}}(\frac{1}{3}-x_{L}(1-x_{L}))]\\
&&m_{T^{-}}=\frac{f}{v}\frac{m_{t}}{\sqrt{x_{L}}}[1+\frac{v^{2}}{f^{2}}(\frac{1}{3}-\frac{1}{2}x_{L}(1-x_{L}))]
\end{eqnarray}
where $\kappa_i$ are the diagonalized Yukawa couplings of the mirror
quarks, $x_{L}$ is the mixing parameter between the SM top-quark $t$
and the new top-quark $T^{+}$.

\section{the description of calculations}
At the tree-level, the Feynman diagrams of the process $pp \to
t\bar{t}$ are shown in Fig.1. The complete one-loop QCD corrections
to the process $pp \to t\bar{t}$ can be generally divided into
several parts: self-energies, vertex corrections, boxes and their
relevant counter terms. If the amplitudes are performed at the order
$\mathcal O(\alpha_{s}^{3})$, we find that the new particles in the
LHT model only can contribute to the self-energies and the relevant
counter terms, which means that there are only the LHT QCD one-loop
virtual corrections to the process $pp \to t\bar{t}$. The relevant
Feynman diagrams for the subprocesses $gg \to t\bar{t}$ and
$q\bar{q} \to t\bar{t}$ in the LHT model are depicted in Fig.2. We
can see that the LHT QCD one-loop virtual corrections
($\Delta\sigma_{\text{LHT}}$) come from the fermion loops. The
one-loop QCD corrected production cross section of the process $pp
\to t\bar{t}$ can be obtained by
\begin{equation}
\sigma_{\text{tot}} = \sigma_{\text{tree}} +
\Delta\sigma_{\text{LHT}}
\end{equation}

In the 't Hooft-Feynman gauge, we use the dimensional reduction
method to regulate the ultraviolet (UV) divergences in the fermion
loops and adopt the on-shell renormalization scheme to remove them.
We list the explicit expressions of these amplitudes in Appendix. We
have checked that the UV divergences in the renormalized propagator
have been canceled. Due to no massless particles in the loop, there
are no infrared (IR) singularities in the one-loop integrals. In our
numerical calculations, we use the parton distribution function
CTEQ10\cite{cteq10} with renormalization/factorization scale $\mu_R
= \mu_F = m_t$.

\begin{figure}[htbp]
\scalebox{0.5}{\epsfig{file=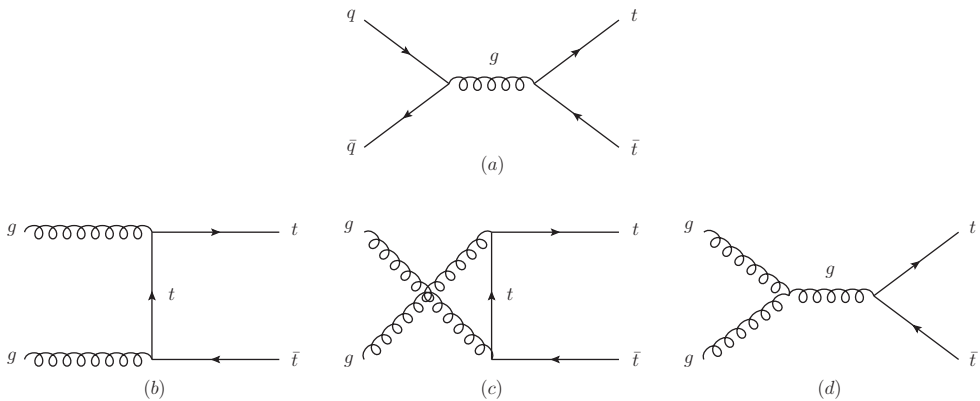}}\vspace{-1cm}\caption{Tree-level
Feynman diagrams of the process $pp\rightarrow t\bar{t}$ in the LHT
model. }
\end{figure}
\begin{figure}[htbp]
\scalebox{0.47}{\epsfig{file=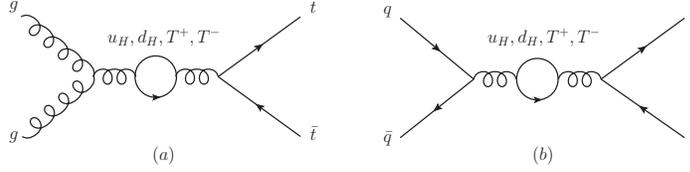}}\vspace{-0.5cm}\caption{Feynman
diagrams of the one-loop QCD correction to the process
$pp\rightarrow t\bar{t}$ in the LHT model. }
\end{figure}

The relevant LHT parameters are the scale $f$, the mixing parameter
$x_{L}$ and the Yukawa couplings $\kappa_{i}$. For the mirror
fermion masses, we get $m_{u_{H}^{i}}=m_{d_{H}^{i}}$ at $\mathcal
O(\upsilon/f)$ and assume that the masses of the first two
generations are degeneracy:
\begin{equation}
m_{u_{H}^{1}}=m_{u_{H}^{2}}=m_{d_{H}^{1}}=m_{d_{H}^{2}},~~~~m_{u_{H}^{3}}=m_{d_{H}^{3}}
\end{equation}

Recently, both of the CMS and ATLAS collaborations reported their
search results of the fermionic top partner and respectively
excluded the masses regions below 557 GeV \cite{CMS-t}and 656
GeV\cite{ATLAS-t} at 95\% CL. In our numerical calculations, we
consider the constraints above and scan the parameter regions:
$f=500\sim2000$GeV, $x_{L}=0.1\sim0.9$,
$\sqrt{2}\kappa_{i}=0.6\sim3$. And we require our samples to satisfy
the constraints from Refs.\cite{constraint}.

We take the input parameters of the SM as\cite{parameters}
\begin{eqnarray} \nonumber
\sin^{2}\theta_{W}=0.231,\alpha_{s}=0.1076,\alpha_{e}=1/128,~~~~~~~~\\
M_{Z}=91.1876\textmd{GeV},m_{t}=173.5\textmd{GeV},m_{h}=125\textmd{GeV}.
\end{eqnarray}

We will discuss the LHT QCD one-loop corrections to the
(un)polarized top pair production by using the following
observables:

\begin{itemize}
\item[(i)] For the unpolarized $t\bar{t}$ production, we calculate the
relative corrections for total $t\bar{t}$ production cross
section($\delta\sigma/\sigma$), which is defined by:
\begin{eqnarray}
&&\delta\sigma/\sigma=\frac{\sigma_{tot}-\sigma_{tree}}{\sigma_{tree}}.
\end{eqnarray}

\item[(ii)] For the polarized $t\bar{t}$ production, we calculate the
spin correlation($\delta C$)\cite{top spin SM}, which is defined by:
\begin{eqnarray}
&&C=\frac{(\sigma_{RR}+\sigma_{LL})-(\sigma_{RL}+\sigma_{LR})}{\sigma_{RR}+\sigma_{LL}+\sigma_{RL}+\sigma_{LR}},\\
&&\delta C=\frac{C_{tot}-C_{tree}}{C_{tree}}.
\end{eqnarray}
Here, the subindices \emph{L(R)} represent left-
($\lambda_{t(\bar{t})}=-1/2$) and
right-handed($\lambda_{t(\bar{t})}=+1/2$) top(antitop) quarks,
respectively.
\end{itemize}
\section{Numerical results and discussions}
\subsection{The correction to the $t\bar{t}$ production cross section}
\begin{figure}[htbp]
\begin{center}
\scalebox{0.2}{\epsfig{file=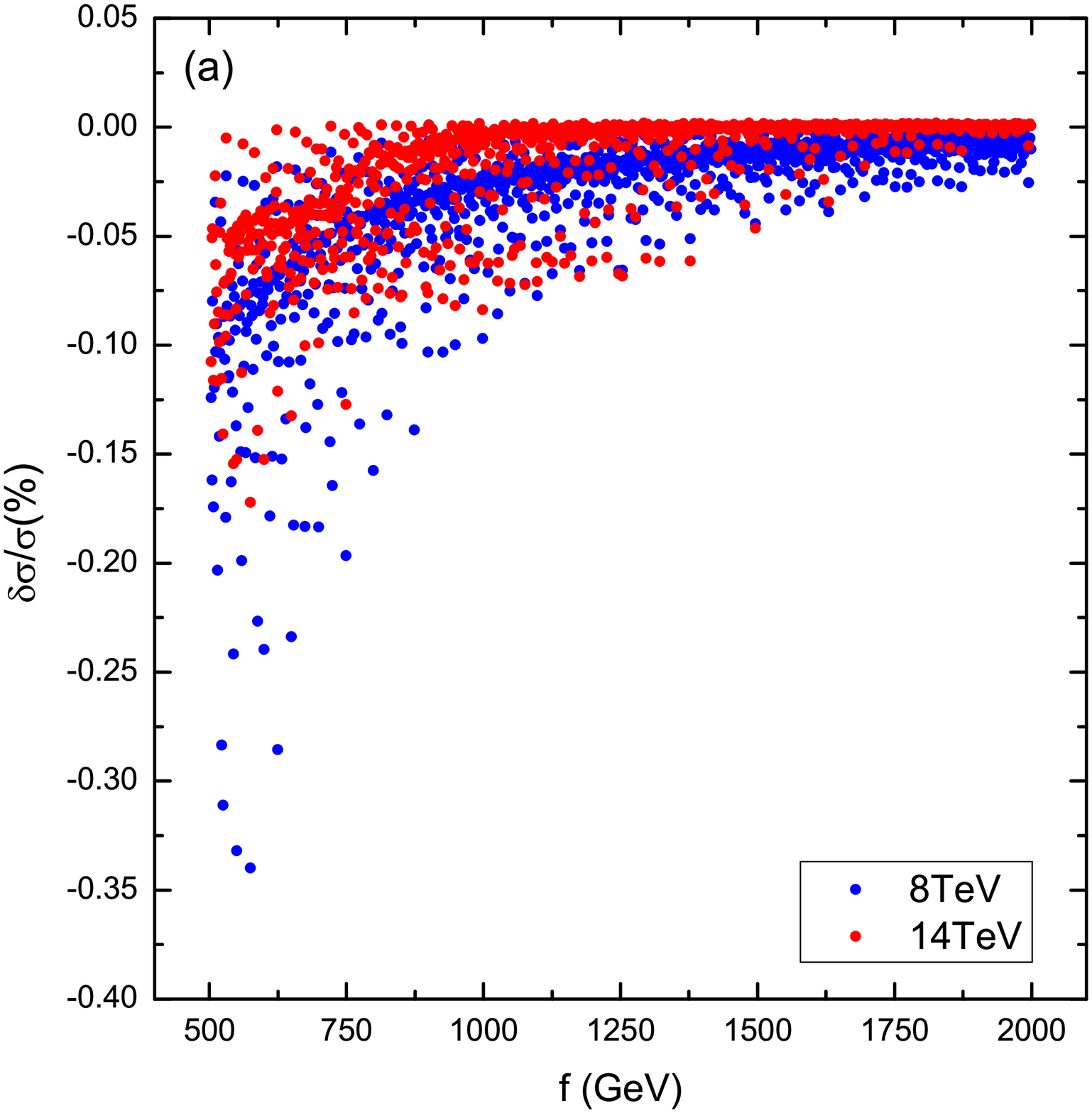}}\hspace{-1cm}
\scalebox{0.2}{\epsfig{file=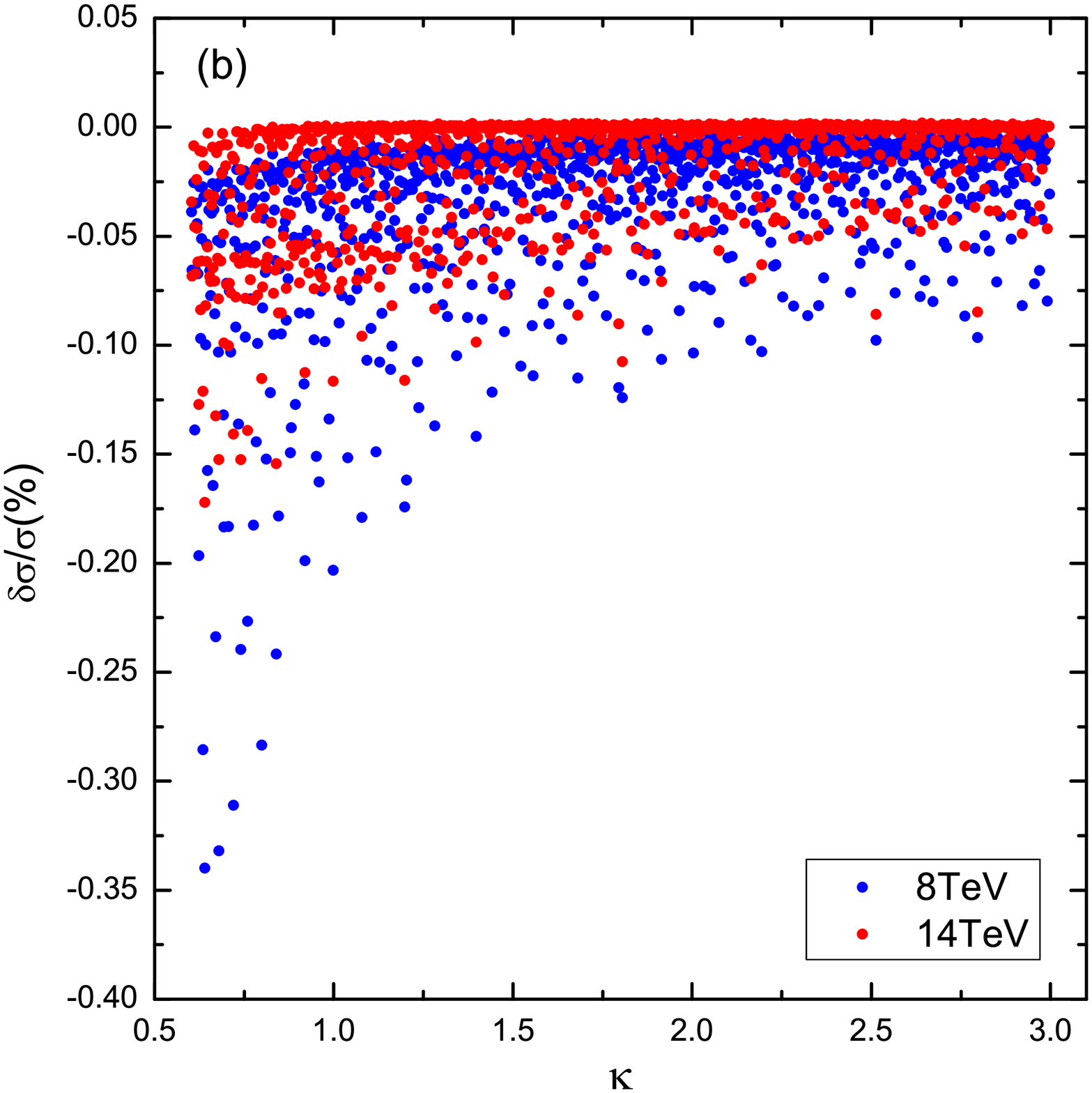}}\hspace{-1cm}
\scalebox{0.2}{\epsfig{file=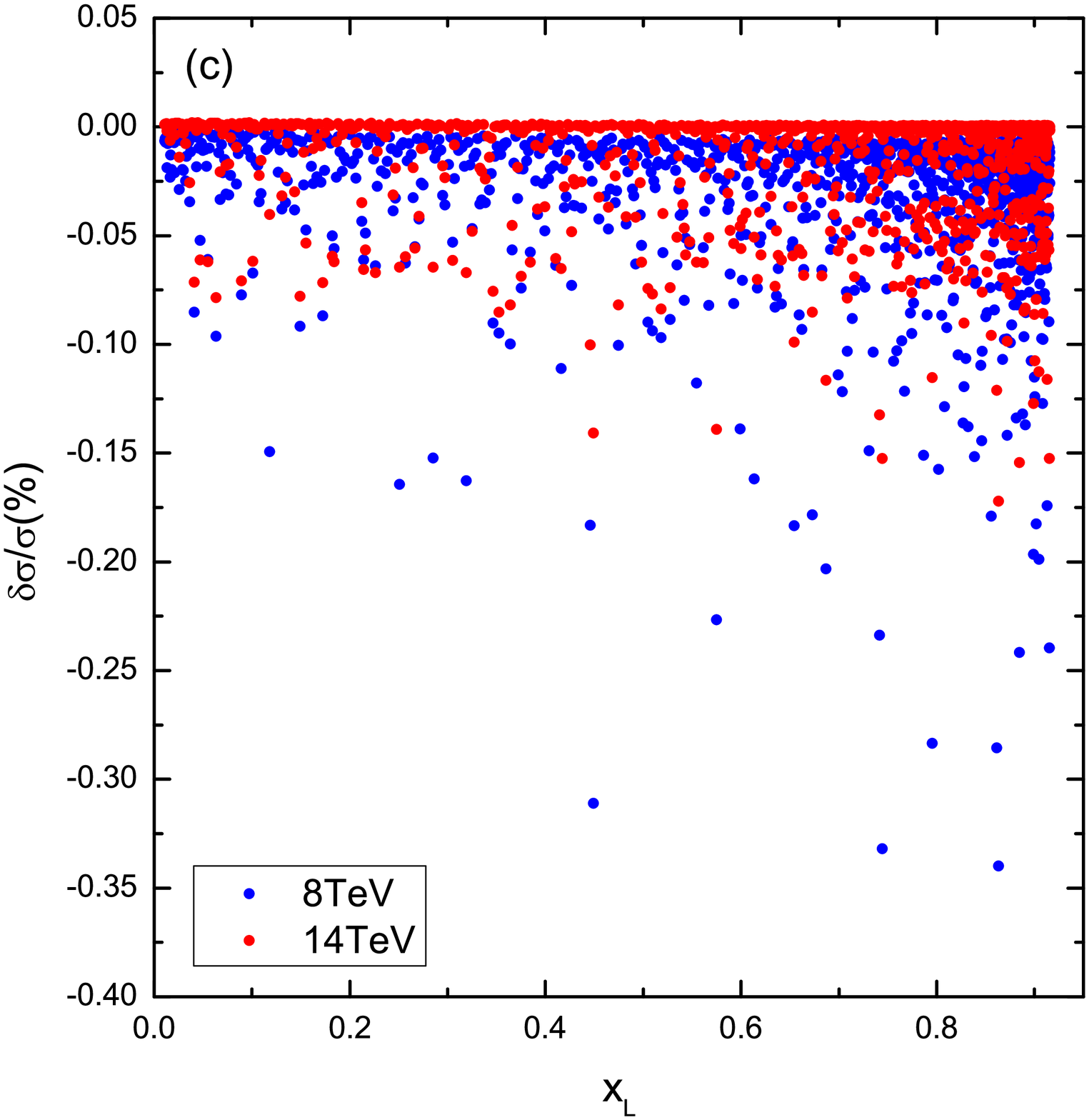}}\hspace{-1cm}\vspace{-1cm}
\caption{The relative correction of $t\bar{t}$ production cross
section $\delta \sigma/\sigma$ as the function of $f,\kappa, x_{L}$
for $\sqrt{s}=8$ TeV and $\sqrt{s}=14$ TeV, respectively.}
\end{center}
\end{figure}

In Fig.3(a), we show the relative correction of the $t\bar{t}$
production cross section $\delta \sigma/\sigma$ as the function of
the scale $f$ at the LHC with $\sqrt{s}=8,14$ TeV respectively. We
can see that the maximum value of the relative correction to the
$t\bar{t}$ production cross section can reach $-0.35$\% for
$\sqrt{s}=8$ TeV and $-0.17$\% for $\sqrt{s}=14$ TeV, respectively.
The results indicate that the size of the corrections and the
sensitivity on these effects are larger at 8 TeV than at 14 TeV.
This is because the correction of the subprocess $gg\rightarrow
t\bar{t}$ is positive and the correction of the subprocess
$q\bar{q}\rightarrow t\bar{t}$ is negative so that they cancel each
other. Due to the main correction comes from $q\bar{q}\rightarrow
t\bar{t}$, the relative correction of the $ t\bar{t}$ total cross
section is negative. When the center-of-mass energy $\sqrt{s}$
varies from 8 TeV to 14 TeV, the correction of the subprocess
$gg\rightarrow t\bar{t}$ increases quickly so that the cancel
between this two subprocesses become stronger. Besides, when the
scale $f$ increases, the relative corrections $\delta \sigma/\sigma$
become small, which indicates that the LHT QCD one-loop effects on
$t\bar{t}$ production cross section will decouple at the high scale
$f$.

In Fig.3(b), we show the relative correction of the $t\bar{t}$
production cross section $\delta \sigma/\sigma$ as the function of
the Yukawa couplings $\kappa$ at the LHC with $\sqrt{s}=8,14$ TeV
respectively. We can see that the relative corrections $\delta
\sigma/\sigma$ decrease with the Yukawa couplings $\kappa$
increasing, which indicates that the LHT QCD one-loop effects on
$t\bar{t}$ production cross section also decouple at the heavy
mirror quark masses.

In Fig.3(c), we show the relative correction of the $t\bar{t}$
production cross section $\delta \sigma/\sigma$ as the function of
the mixing parameter $x_{L}$ at the LHC with $\sqrt{s}=8,14$ TeV
respectively. The distribution behaviors of the samples can be
explained as follows. According to the Eqs.(4, 5), we can see that
the heavy top quark $T^{+}$ and $T^{-}$ masses have a strong
dependence on the mixing parameter $x_{L}$. When $x_{L}\to 0$, the
masses of $T^{+}$ and $T^{-}$ will become heavy and their
contribution become very small. When $x_{L}\to 1$, the masses of
$T^{+}$ will become heavy while the masses of $T^{-}$ will become
light. As a result, the effect of $T^{-}$ will still reside in the
$t\bar{t}$ production.

\subsection{The correction to the $t\bar{t}$ spin correlation}

\begin{figure}[htbp]
\begin{center}
\scalebox{0.2}{\epsfig{file=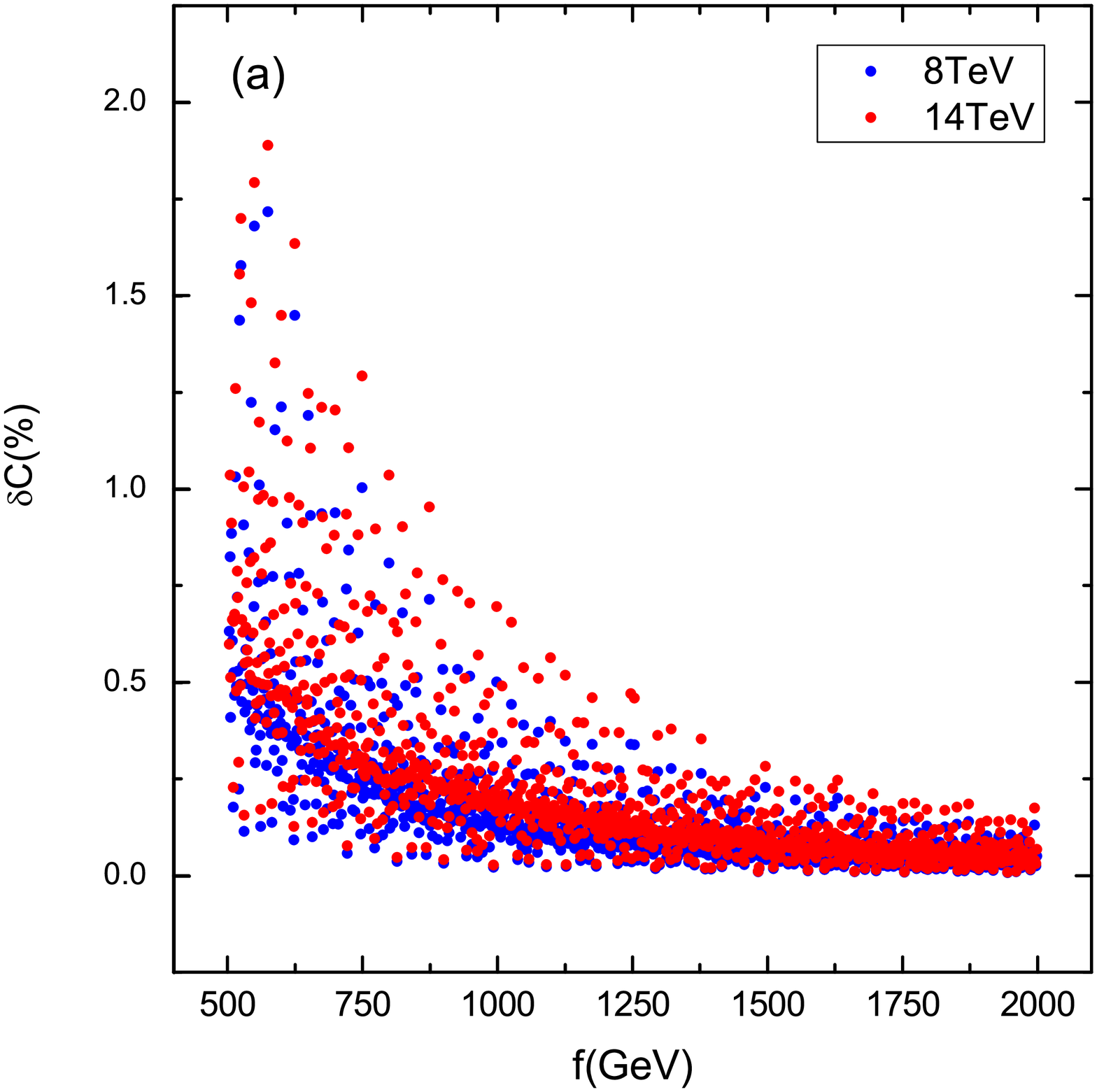}}\hspace{-1cm}
\scalebox{0.2}{\epsfig{file=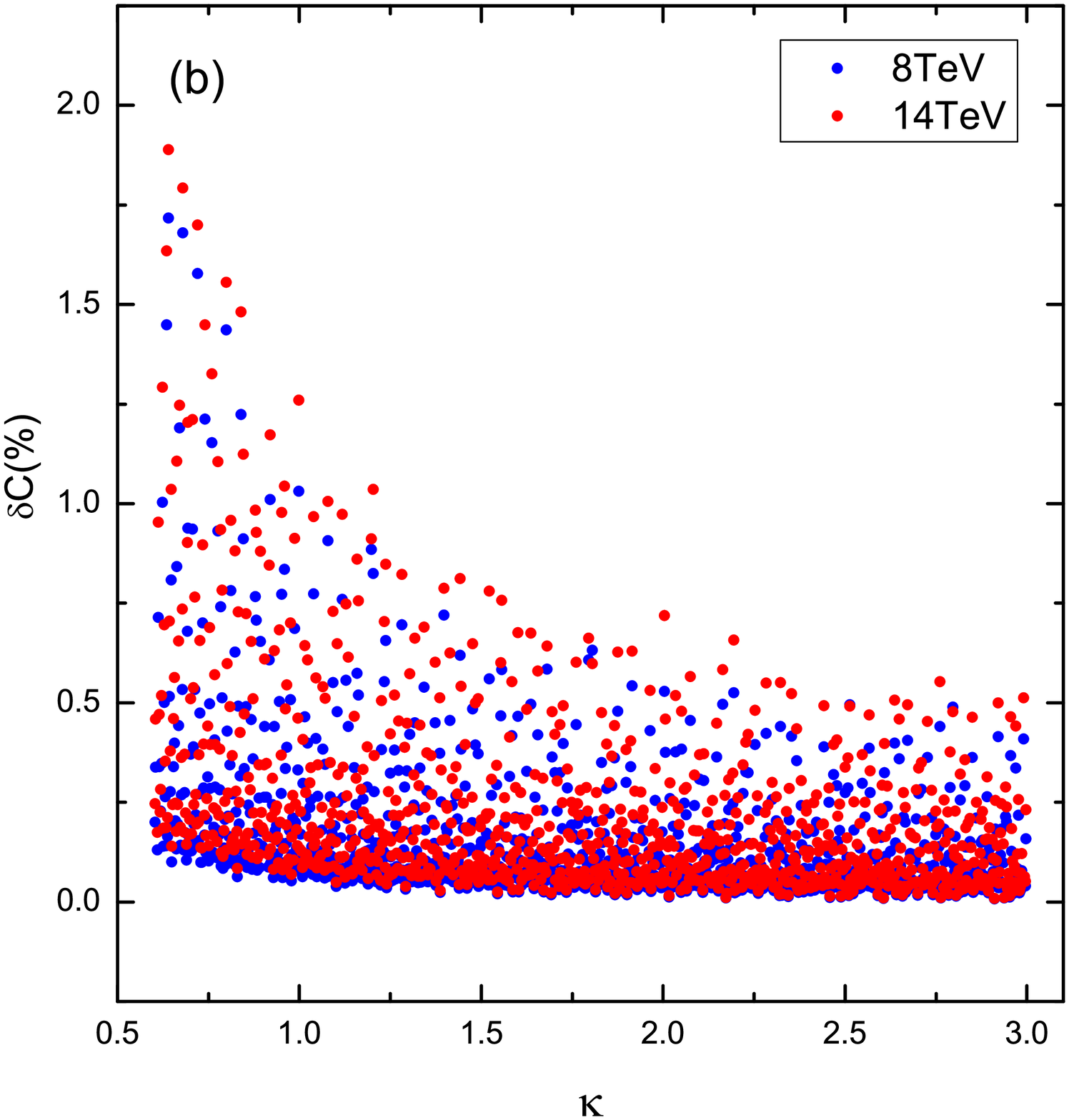}}\hspace{-1cm}
\scalebox{0.2}{\epsfig{file=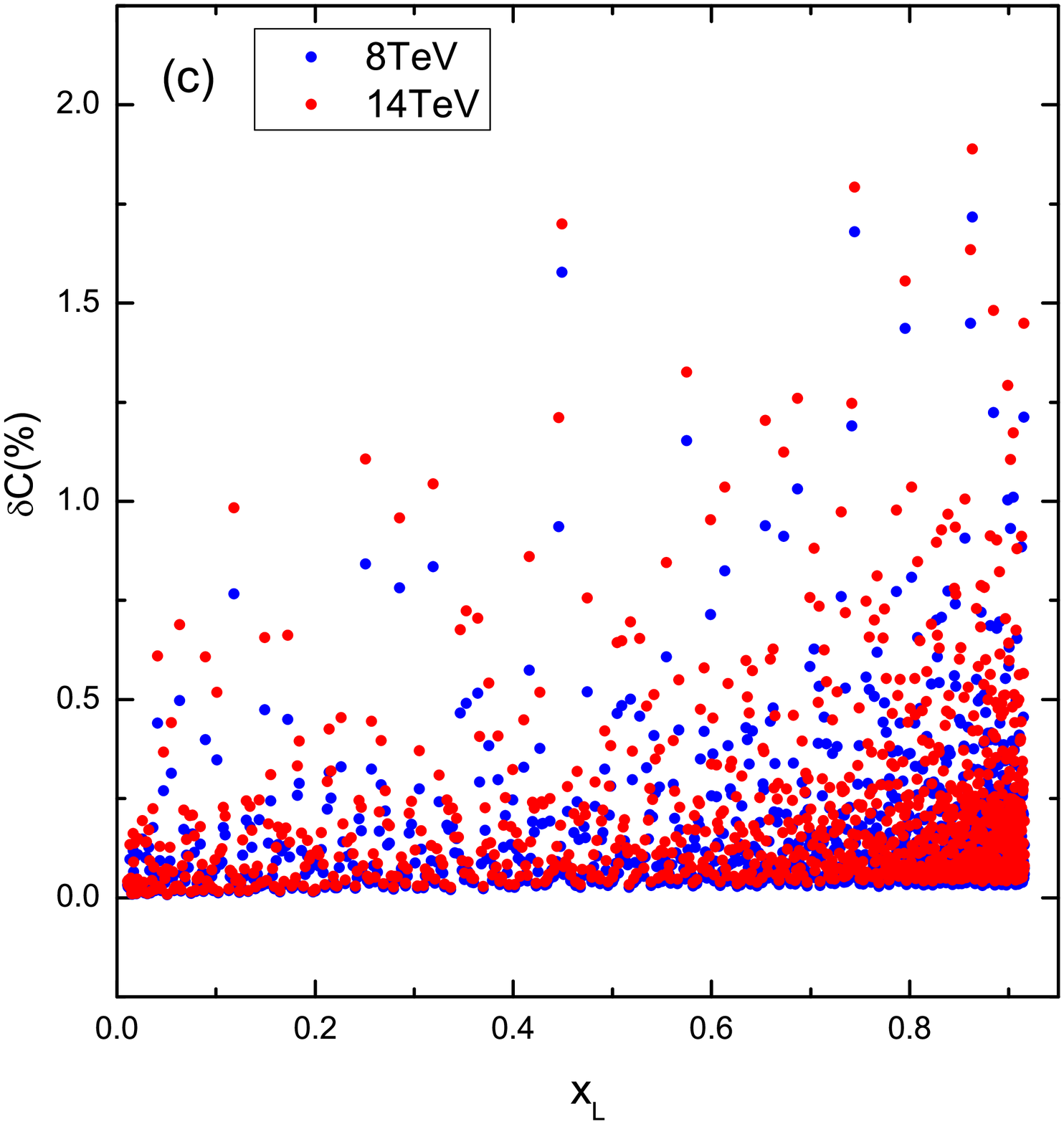}}\hspace{-1cm}\vspace{-1cm}
\caption{The relative correction of the $t\bar{t}$ spin correlation
$\delta C$ as the function of $f,\kappa, x_{L}$ for $\sqrt{s}=8$ TeV
and $\sqrt{s}=14$ TeV, respectively.}
\end{center}
\end{figure}

Recently, the CMS collaboration reported their measurement of the
$t\bar{t}$ spin correlation coefficient $C=0.24\pm 0.02(stat.)\pm
0.08(syst.)$ in the helicity basis\cite{dcLHC2}, which is agree with
the SM predictions. In Fig.4, we show the relative correction of the
$t\bar{t}$ spin correlation $\delta C$ as the function of $f,\kappa,
x_{L}$ for the LHC with $\sqrt{s}=8,14$ TeV, respectively. We can
see that the behaviors of $\delta C$ versus $f,\kappa, x_{L}$ are
similar to those of the relative correction $\delta \sigma/\sigma$.
The maximum value of $\delta C$ can reach about $1.7\%$ for
$\sqrt{s}=8$ TeV and $2\%$ for $\sqrt{s}=14$ TeV, which may be
detected at the LHC\cite{dcLHC1}.

\section{summary} \noindent

In this paper, we studied the one-loop $\mathcal O(\alpha_{s}^{3})$
QCD corrections to $t\bar{t}$ production in the LHT model at the LHC
for $\sqrt{s}=8,14$ TeV. We presented the numerical results for the
relative correction to $t\bar{t}$ production cross section and
$t\bar{t}$ spin correlation at the LHC. The relative correction of
the $t\bar{t}$ production cross section is negative and only can
reach $-0.35\%$. The relative correction of the $t\bar{t}$ spin
correlation $\delta C$ can reach about $1.7\%$ for $\sqrt{s}=8$ TeV
and $2\%$ for $\sqrt{s}=14$ TeV, which may be a potential probe to
detect the LHT effects at the LHC.

\textbf{Acknowledgments}

This work is supported by the National Natural Science Foundation of
China under grant Nos.11347140, 11305049, 11405047 and the China
Postdoctoral Science Foundation under grant No. 2014M561987.

\begin{center}
\textbf{Appendix: The explicit expressions of the renormalized gluon
propagator} \cite{renormalization}
\end{center}

They can be represented in form of 1-point and  2-point standard
functions $A,B_{0},B_{1}$. Here $p_{t}$ and $p'_{t}$ denote the
momenta of the top and antitop respectively, and they are assumed to
be outgoing.

Renormalization gluon propagator
\begin{figure}[htbp]
\scalebox{0.45}{\epsfig{file=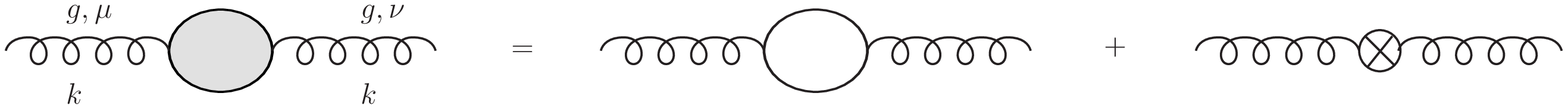}}\vspace{-0.5cm}
\end{figure}
\begin{eqnarray*}
-i\hat{\Sigma}_{\mu\nu}(k)=-i\Sigma_{\mu\nu}(k)+(-i\delta\Sigma_{\mu\nu}(k))
\end{eqnarray*}
where
\begin{eqnarray*}
&&\Sigma_{\mu\nu}(k)=g_{\mu\nu}\Sigma_{T}(k)+k_{\mu}k_{\nu}\Sigma_{L}(k)\\
&&\delta\Sigma^{gg}_{\mu\nu}(k)=g_{\mu\nu}[\delta Z_{gg}k^{2}]\\
&&\delta Z_{gg}=-\frac{\partial\Sigma^{gg}_{T}(k^{2})}{\partial
k^{2}}|_{k^{2}=0}\\
&&\hat{\Sigma}^{gg}_{\mu\nu}=\hat{\Sigma}^{gg}_{\mu\nu}(u_{H})+\hat{\Sigma}^{gg}_{\mu\nu}(d_{H})+\hat{\Sigma}^{gg}_{\mu\nu}(T^{\pm})
\end{eqnarray*}
\begin{eqnarray}
&&-i\hat{\Sigma}_{\mu\nu}^{gg}(k)=\frac{-4ig_{s}^{2}g_{\mu\nu}T_{\alpha\beta}^{a}T_{\beta\alpha}^{b}}{16\pi^{2}}\nonumber\\
&&\{[-\frac{2}{3}A_{0}+\frac{2}{3}m^{2}_{T^{-}}B_{0}-\frac{2}{3}k^{2}B_{1}(k,m_{T^{-}},m_{T^{-}})+\frac{2}{3}m^{2}_{T^{-}}-\frac{k^{2}}{9}]\nonumber\\
&&+[-\frac{2}{3}A_{0}+\frac{2}{3}m^{2}_{T^{+}}B_{0}-\frac{2}{3}k^{2}B_{1}(k,m_{T^{+}},m_{T^{+}})+\frac{2}{3}m^{2}_{T^{+}}-\frac{k^{2}}{9}]\nonumber\\
&&+[-\frac{2}{3}A_{0}+\frac{2}{3}m^{2}_{u_{H}}B_{0}-\frac{2}{3}k^{2}B_{1}(k,m_{u_{H}},m_{u_{H}})+\frac{2}{3}m^{2}_{u_{H}}-\frac{k^{2}}{9}]\nonumber\\
&&+[-\frac{2}{3}A_{0}+\frac{2}{3}m^{2}_{d_{H}}B_{0}-\frac{2}{3}k^{2}B_{1}(k,m_{d_{H}},m_{d_{H}})+\frac{2}{3}m^{2}_{d_{H}}-\frac{k^{2}}{9}]\nonumber\\
&&-[\frac{2}{3}m^{2}_{T^{-}}\frac{\partial B_{0}}{\partial
k^{2}}-\frac{2}{3}B_{1}(0,m_{T^{-}},m_{T^{-}})-\frac{1}{9}]k^{2}\nonumber\\
&&-[\frac{2}{3}m^{2}_{T^{+}}\frac{\partial B_{0}}{\partial
k^{2}}-\frac{2}{3}B_{1}(0,m_{T^{+}},m_{T^{+}})-\frac{1}{9}]k^{2}\nonumber\\
&&-[\frac{2}{3}m^{2}_{u_{H}}\frac{\partial B_{0}}{\partial
k^{2}}-\frac{2}{3}B_{1}(0,m_{u_{H}},m_{u_{H}})-\frac{1}{9}]k^{2}\nonumber\\
&&-[\frac{2}{3}m^{2}_{d_{H}}\frac{\partial B_{0}}{\partial
k^{2}}-\frac{2}{3}B_{1}(0,m_{d_{H}},m_{d_{H}})-\frac{1}{9}]k^{2}\}
\end{eqnarray}

\end{document}